# NUMERIC MODELING OF FIRE SUPPRESSION BY ORGANOPHOSPHOROUS INHIBITORS

**G. M. Makhviladze[2], S. E. Yakush[1]*, A.P. Zykov[1]***

1*Institute for Problems in Mechanics, Russian Academy of Sciences, Moscow, 119526, Ave. Vernadskogo 101, Russia*

2*University of Central Lancashire, Preston, PR1 2HE, UK*

**Corresponding author: zykov@ipmnet.ru , yakush@ipmnet.ru*

## ABSTRACT

*Numerical calculations of the effect of organophosphorous inhibitor $(CF_3CH_2O)_3P$ and its mixtures with carbon dioxide on propane flames are carried out using the three dimensional Reynolds-averaged Navier-Stokes (RANS) equations in the low Mach number approximation. The k-e model of turbulence, the EDC combustion model and the weighted-sum-of-gray-gases model of radiation are used. The Westbrook global-kinetic scheme with fractional order of reaction was used for the calculation of chemical reaction rate of propane combustion. The empirical expression for the correction factor for the chemical reaction rate was used to model the effect of organophosphorous inhibitor no the reaction. Two series of test calculations for different values of the correction factor are carried out. Dependences of the minimum extinguishing concentration of the inhibitor per carbon dioxide volume concentration in the extinguishing mixtures were obtained. The results of test calculations are shown to agree reasonably with the experimental data. A calculation of the compartment fire extinguishment was carried out using the result of test calculations. Temperature and inhibitor volume concentration fields at the moment of fire extinguishment are obtained. The results of calculation are used to find out the optimum position of the extinguish mixture source.*

## Introduction

Numerical modeling, as one of the most effective methods for studying complex physical processes, is used with increasing frequency for studying the chemically reactive gas flows, including the efficiency of extinguishing means (reactive gases, powders, sprinklers) [1]. Particularly, it is belong in full measure to studying compartment fire and seeking more effective fire suppressants. Use of extinguish mixture with organophosphorous inhibitors is one of the more perspective line of fire suppressant developments. High boiling-point and low volatility conditions using aerosol technology for fire suppressants delivery to the fire source. However, efficiency of organophosphorous fire suppressant not depends on methods of its delivery to fire source, as was shown in [2]. Thus, combustion suppression effect can be taken into consideration by means of a correction factor for chemical reaction rate, and this fact simplify a numerical modeling of its influence on fire substantially.

Numerical modeling of the effects of organophosphorous inhibitor $(CF_3CH_2O)_3P$ and its mixtures with carbon dioxide on propane combustion are carried out in this work. In the first part the test calculations of inhibitor influence on fire were carried out. Calculation domain was chosen in accordance with the experimental data in works [3, 4], which were obtained by means of “cup-shaped burner”. Domain sizes were chosen 0.5×0.5×1.0 m. Fuel source with the sizes 0.02×0.02 m was positioned on the “cup-shaped burner”, which sizes were chosen 0.06×0.06 m and it was located at the center of bottom border. Other part of bottom border was free and air with extinguish mixture were coming up through it. Top border of domain was free and flank border were adiabatic wall. The empirical expression for the correction factor for the chemical reaction rate was obtained in works [3, 4]. Testing of this model for the effect of the organophosphorous inhibitor on the chemical reaction is one of the major purposes of this work. Three series may are singled out from the results of 57 tests calculations, which are presented below. Two series were carried out with different values of the parameter in the model for inhibitor influence on the reaction. Time dependences of heat release rate and maximum temperature in the domain, the temperature and extinguishing mixture component concentration distributions are obtained during the calculations. The minimum extinguishing concentration of inhibitor was obtained for each chosen value of carbon dioxide concentration in extinguishing mixture.

The compartment fire extinguishment was calculated using the model of organophosphorous inhibitor effect on the reaction rate in the second part of this work. Time dependence of heat release rate, the temperature and inhibitor concentration, spatial distributions at the different times are obtained during this calculation.

## Mathematical method

Substantially subsonic single-phase compressible gas flows in a compartment with an opening are considered. The mathematical model is based on the Reynolds-Averaged Navier-Stokes (RANS) equations describing three-dimension mass, momentum and energy conservation in the Euler coordinates. The low Mach number model [5] is used in which the dynamic gas compressibility is neglected, but the gas density dependence on the average pressure and temperature are taken into account. The total pressure is represented as the sum of the domain-average value Po and the dynamic pressure: P = Po + p. The equation of state contains only the average pressure:

$$\rho = \frac{P_o}{R_g T} \sum_{\alpha} \frac{Y^{\alpha}}{M^{\alpha}} \tag{1}$$

where $\rho$ is the gas density, $R_g$ is the universal gas constant, $T$ is the temperature, $Y^{\alpha}$ and $M^{\alpha}$ are the mass fraction and molar mass of the $\alpha$-th species. The pressure deviation p from the average level ("dynamic" pressure) is considered only in the gradient term of the momentum equation. This assumption eliminates the effects associated with the acoustic waves and allows an effective numerical algorithm to be used.

Table 1

| Equation | $\Phi$ | $\Gamma_{eff}^{\Phi}$ | $S_{\Phi}$ |
|---|---|---|---|
| Continuity | 1 | 0 | 0 |
| Navier-Stokes | $u_i$ | $\mu_{eff}$ | $-\frac{\partial p}{\partial x_i} + \frac{\partial}{\partial x_j}\left(\Gamma_{eff}^{u} \frac{\partial u_j}{\partial x_i}\right) - \frac{2}{3}\frac{\partial}{\partial x_i}\left(\Gamma_{eff}^{u} \frac{\partial u_j}{\partial x_j}\right) - \rho g \delta_{i3}$ |
| Enthalpy | $h$ | $\mu_{eff}/Pr$ | $\sum_{\alpha} R^{\alpha} h_{\alpha}^{o} + \frac{\partial q_i^r}{\partial x_i}$ |
| Concentrations | $Y^{\alpha}$ | $\mu_{eff}/Sh^{Y}$ | $R^{\alpha}$ |
| Turbulence energy | $k$ | $\mu_{eff}/Sh^{k}$ | $-\rho\varepsilon + G_k + G_b$ |
| Dissipation rate | $\varepsilon$ | $\mu_{eff}/Sh^{e}$ | $\frac{\varepsilon}{k}\left[C_1^t (G_k + G_b) - C_2^t \rho\varepsilon\right]$ |

$$G_k = 2\mu_t \left[\sum_i \left(\frac{\partial u_i}{\partial x_i}\right)^2\right] + \mu_t \left[\sum_{i>j}\left(\frac{\partial u_i}{\partial x_j} + \frac{\partial u_j}{\partial x_i}\right)^2\right] \qquad G_b = \mu_t g \frac{1}{\rho}\frac{\partial \rho}{\partial x_3}$$

$$\mu_{eff} = \mu_l + \mu_t \qquad \mu_t = C_{\mu} \frac{\rho k^2}{\varepsilon}$$

*Governing equations*

The gas density, velocity, dynamic pressure, gas species concentrations, specific enthalpy, kinetic energy of turbulence and its dissipation rate are used as the basic variables. The turbulent viscosity is calculated by means of the standard k- e model of turbulence.

All transfer equations may be written as

$$\frac{\partial}{\partial t}(\rho\Phi)+\frac{\partial}{\partial x_i}(\rho u_i\Phi)=\frac{\partial}{\partial x_i}\left(\Gamma_{eff}^{\Phi}\frac{\partial\Phi}{\partial x_i}\right)+S_{\Phi}\ , \qquad (2)$$

and source members are represented in Table 1. Standard values of the turbulent model constants, turbulent Schmidt and Prandtl numbers are used [1].

The dependence of specific heat on temperature is taken into account through its Taylor series decay by temperature:

$$c_p=\sum_{\alpha}Y^{\alpha}c_p^{\alpha}=\sum_{\alpha}Y^{\alpha}\sum_{k=0}^{4}A_{\alpha}^{k}T^{k} \qquad (3)$$

where Taylor coefficients for α-species $A_{\alpha}^{k}$ is a table values and there can be found in [6] for example.

*Combustion model*

It is necessary to define turbulent combustion rate for closing the combined equations (1)-(3) which depends not only on the chemical reactions, but also on turbulent mixing. The Eddy Dissipation Concept (EDC) is used in this work to calculate the reaction rate in the turbulent flow [7]. In accordance with this model, two subregions are distinguished in the flow, i.e., the fine structures and the surrounding fluid. The model assumes that chemical reactions proceed in the fine structures only, i.e. in the subregion with the smallest turbulence scales. The two subregions exchange by mass and heat with each other. In the fine structures, the gas species concentrations, density and temperature can be different from their average values. So, the macroscopic variables of fine structures are defined and these will marked with asterisk below. The turbulent characteristics determine origin, mean residence time of fine structures and therein occupied gas mass fraction entirely. The mean residence time of fine structures is defined as:

$$\tau^{*}=0.412\sqrt{\frac{\nu}{\varepsilon}} \qquad (4)$$

where $\nu$ is the kinematic viscosity. The mass fraction $\gamma^*$ occupied by the fine structures is defined by the expression:

$$\gamma^* = \left[2.13\sqrt[4]{\frac{\nu\varepsilon}{k^2}}\right]^2 \qquad (5)$$

The reaction rates $R_\alpha^*$ for the species $\alpha$, which is the mean net mass transfer rate of the species $\alpha$ between the fine structures and surrounding fluid, and the enthalpy $h^*$ in the fine structures are calculated using the mass and energy balances for the well-stirred reactor. These variables are defined by the following equations:

$$R_\alpha^* = \frac{\rho\chi\gamma^*}{\tau^*\left(1-\gamma^*\chi\right)}\left(Y^\alpha - Y_\alpha^*\right) \quad , \qquad (6)$$

$$\frac{\rho^*}{\tau^*\left(1-\gamma^*\chi\right)}\left(Y_\alpha^* - Y^\alpha\right) = R_\alpha \quad , \qquad (7)$$

$$\frac{\rho^*}{\tau^*\left(1-\gamma^*\chi\right)}\sum_\alpha\left(Y_\alpha^* h_\alpha^* - Y^\alpha h_\alpha\right) = \frac{\partial q_i^r}{\partial x_i} \ , \qquad (8)$$

where $\chi$ is correction coefficient indicative reactionable gas mass fraction in fine structure, in all calculations $\chi = 1$; $\partial q_i^r / \partial x_i$ is radiation heat transfer. $R_\alpha$ is summary rate for species $\alpha$ in all reactions which are under consideration:

$$R_\alpha = \sum_{i=1}^{N_r} S_{\alpha,i} R_i \qquad (9)$$

$N_r$ is number of chemical reaction, $S_{\alpha,i}$ is stoichiometric coefficient for species $\alpha$ in $i$-th chemical reaction, $R_i$ is $i$-th chemical reaction rate.

Propane combustion is described by a single gross-reaction

$$C_3H_8 + 5\,O_2 \rightarrow 3\,CO_2 + 4\,H_2O$$

with the global kinetic, which was described in [8, 9]:

$$W_o = A\rho^{\alpha+\beta}\frac{Y_F^\alpha Y_O^\beta}{M_F^\alpha M_O^\beta}\exp\left(-\frac{E_a}{RT}\right) , \qquad (10)$$

where $\alpha$ = 0,1, $\beta$ = 1,65, $M_F$, $M_O$ are propane and oxygen molar masses, $Y_F$, $Y_O$ are propane and oxygen mass fractions, $\rho$ is gas density, $R_g$ is the universal gas constant, $T$ is the temperature, $E_a$ = 126 600 (J) is activation energy, $A = 8{,}6\times10^9$ $(m^3/mol)^{0,75}/s$ is pre-exponential factor.

An empirical expression for the correction factor for describing the effect of organophosphorous inhibitor on the chemical reaction rate was used in modeling on combustion rate. This expression was obtained in works [3, 4], where the program PREMIX was used for methane combustion:

$$K = \left(1 - \xi\sqrt{X^i}\right)^2 , \qquad (11)$$

where $X^i$ is the inhibitor volume fraction in the gas mixture, $\xi$ is an empirical coefficient.

Thus, by solving equations (4)-(8), (10) for the perfectly-stirred reactor the effect of the inhibitor was taken into account: the chemical reaction rate (10) was multiplied by correction factor (11), and the resultant reaction rate for species $\alpha$ was defined by expression instead of (9):

$$R_\alpha = S_\alpha \frac{M_F}{M_\alpha} K \cdot W_o \qquad (12)$$

It is supposing in this work the combustion powerful is low for using domains and propane burn down without soot formation.

***Radiation***

The main equations for radiation heat transfer can be found in works [10, 11]. To calculate the heat loss due to radiation, it is necessary to solve the following equation:

$$\frac{\partial q_i^r}{\partial x_i} = -4K_{abs}\left(\sigma T^4 - E\right) , \qquad (13)$$

where $\vec{q}$ is the radiative energy flux; $K_{abs}$ is the absorption coefficient; $\sigma$ is Stefan-Boltzmann constant; $E$ is radiative energy density.

Absorption coefficient is the optical properties of surrounding gas, and it is strong depended on emission wave length. The optical properties of surrounding gas are described most fully by spectral models, which take into account the absorption coefficient dependences of wave length in full band. But this models demand big computational consumption. Therefore, the weighted-sum-of-gray-gases model is using in this work, which detailed describe can be found in [12] for example. Due to this model the absorption coefficient is calculated by following expression:

$$K_{abs} = \sum_{i_s=1}^{2}\sum_{i_g=0}^{3} a_{g,i_g} a_{s,i_s} \left(k_{g,i_g}\left(P_{CO_2} + P_{H_2O}\right) + k_{s,i_s} f_V\right) \qquad (14)$$

where $k_{\alpha,i_\alpha}$ is the absorption coefficient for model grey gases, and the combined radiant emittance of all this gases approximate the real gas emissivity; $P_{CO_2}$ and $P_{H_2O}$ is carbon dioxide and water vapor partial pressure; $f_V$ is volume fraction of soot. Absorption weight coefficients of grey gases are approximated by polynomials with three degree of the temperature:

$$a_{\alpha,i_\alpha} = \sum_{j=0}^{3} b_{i_\alpha,j}^{\alpha} T^j \quad , \alpha = \text{g, s} \tag{15}$$

Polynomial coefficients $b_{i_\alpha,j}^{\alpha}$ and absorption coefficients for grey gases are the table variables and their values can are found in [12].

The so-called fluxes method is used for solving radiative energy transport equation (13), which is described for example in [13]. The radiative energy density for single direction $E_i$ ( $i=x,y,z$ ) is taken into use toward this purpose, moreover total radiative energy density is defined as

$$E = 1/3 \sum_i E_i \ .$$

Integration (13) for single direction leads to equation

$$\frac{\partial}{\partial x_i} \frac{1}{mK_{abs}} \frac{\partial E_i}{\partial x_i} + mK_{abs}\left(\sigma T^4 - E_i\right) = 0 \tag{16}$$

with boundary conditions

$$\pm \frac{4}{3mK_{abs}} \frac{\partial E_i}{\partial x_i} = \frac{2\varepsilon_w}{2-\varepsilon_w}\left[\frac{2}{3m}\left(E_i + \left(\frac{3}{4m} - \frac{1}{2}\right)\left(E_j + E_k\right)\right) - \sigma T^4\right] \tag{17}$$

where $m \approx 1{,}2$ is the integration constant, $\varepsilon_w = 0{,}8$ is the wall emission coefficient.

***Initial and boundary conditions***

The horizontal components of velocity are equal to zero and vertical component is equal to velocity of the incoming into domain gas at the beginning of test calculations. Gas is quiescent in the entire domain for the calculation of compartment fire extinguishment. The pressure distribution corresponds to the hydrostatic equilibrium, and all other variables have the same constant values in the whole computational domain. The walls are considered to be no-slip and adiabatic. The logarithmic law for estimation of near-wall turbulent characteristics and shear stresses are implemented in the wall functions [14]. The Neumann conditions are posed on the solid walls for the pressure, enthalpy and gas species concentrations. On the free boundary, the normal derivative of all velocity components is set to zero, the pressure is equal to the ambient pressure (i.e. the dynamic pressure is zero). For the outflow parts of free boundaries, the Neumann conditions are used for the enthalpy, gas species concentrations and turbulent quantities. For the inflow parts all these variables are set to their ambient values. The fire source is modeled by a burner with a given mass flowrate of fuel. It is assumed that the total area of burner holes is small, so that the dynamic pressure meets the Neumann conditions, the velocity components tangential to the inlet boundary are set to zero. Ignition of the fuel entering the compartment through the burner occurs by a heat source in the cells adjacent to the burner.

***Numerical method***

The numerical solution of fluid dynamic combined equations (1)-(3) is based on implicit SIMPLE method [15], and finite-difference numerical transport equations are solved on the 3D MAC-grid. Due to this method the correction of pressure and velocity fields for the purpose of satisfying continuity equation was carried out after preliminary velocity calculations with using pressure from old iteration. Multigrid method [16] is used for solving the dynamic pressure correction Poisson equation at the each step of global iterations. This method allows achieving high rate of convergence regardless of grid size and irregularity. All transport equations are solving by 3D TDMA method (look in [15] for example). Linear interpolation is using for calculate variables value in space points their not containing. Upstream differences scheme is using for monotonic solution achieved by modeling space differentials. Scheme with second order of accuracy is using for modeling convective and viscous terms.

The main part of the code is incarnated by means of C++ programming language using object-oriented approach. The subroutines for multigrid method and calculation radiant heat loss by means of equation (14)-(17) are written in FORTRAN-90.

## Result of test calculations

All test calculations were performed on a 30×30×35 mesh with minimum and maximum length 0.01 m and 0.06 m correspondingly and with a variable time step. The flame heat release and maximum temperature in the compartment were defined, the velocity, temperature, mass fraction of carbon dioxide and inhibitor fields in the vertical symmetry plane were built during the calculations. At the first stage the calculation was carried out until the steady-state combustion was obtained on conditions that pure air coming up through the bottom boundary. Fuel consumption was equal $2 \cdot 10^{-2}$ g/s which ensure heat release 1.3 kW at the normal conditions. Fields distributions were obtained at that are using as start data for subsequent calculations with introduction from below the extinguishing mixture with given volume fraction. The gas velocity on the bottom boundary and the fuel consumption were keeping constants at the same time. The ambient temperature was equal to $20^{o}$C, the gas velocity on the bottom boundary was equal 0.13 m/s.

The flame heat release and the maximum temperature receded to a greater or lesser extent against the extinguish mixture compound and concentration since its coming up into domain. Time dependences of observable variables gets strongly oscillating view and burn could be terminating. The same flows movies are observed in all calculations with the only difference that the burn can be terminating or keeping against of extinguish mixture compound and concentration. As it was agree in work [3, 4] the extinguishing mixture provide burn termination on the assumption of chemical reaction was break up in 10 seconds after the beginning of its coming up. For example, the heat release and the maximum temperature time dependences for two different cases are presented on Fig. 1 (calculation with burn termination) and Fig. 2 (without burn termination). At the some calculations the short term fall of heat release to some tens watt was observed, but the combustion was keep up and then heat release run back, as it can see on Fig. 2.

Also, considerable distinctions on temperature and concentrations distributions for various calculations were not observed. Unperturbed before extinguish mixture coming up into domain the flame convective plume collapsed little by little and began to revolve, as that can be seen on Fig. 3. The extinguishing mixture hit upon combustion region as soon as entered into domain, and its concentration achieved to set point in all domain gradually with the exception of small regions on the top free border, where the cold air permeated from the outside. This small heterogeneity of concentrations on the top border not influenced to the calculations results because of enough remoteness free border from the fire and permanent gas entry through the bottom border. Distributions of carbon dioxide mass concentration fields for one of the calculations are presented on Fig. 4.

Table 2. Flame extinguishment times for calculations with carbon dioxide.

| $CO_2$ | 20 | 25 | 26 | 27 | 30 | 40 | 50 |
|---|---|---|---|---|---|---|---|
| *t* | × | × | 9,8 | 7 | 4 | 2,6 | 0,8 |

a) b)

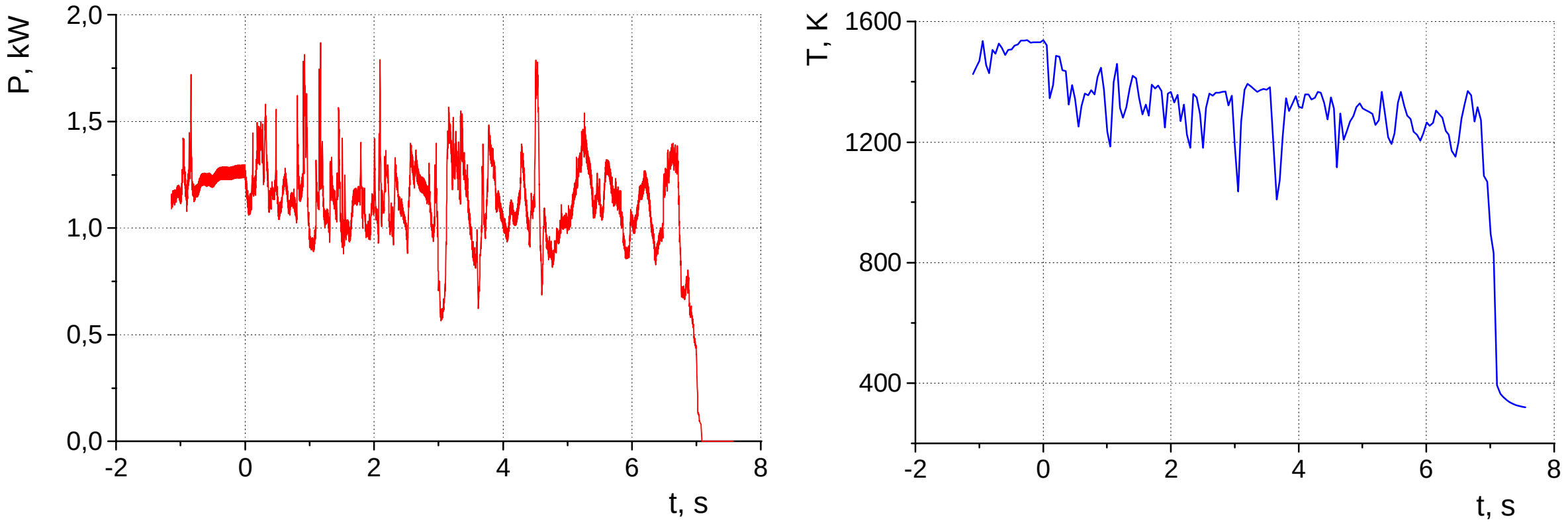

Fig. 1. Flame heat release (a) and maximum temperature in domain (b) at the calculation with extinguish mixture compound $X_{CO_2} = 27\%$ , $X_{inh} = 0\%$.

a) b)

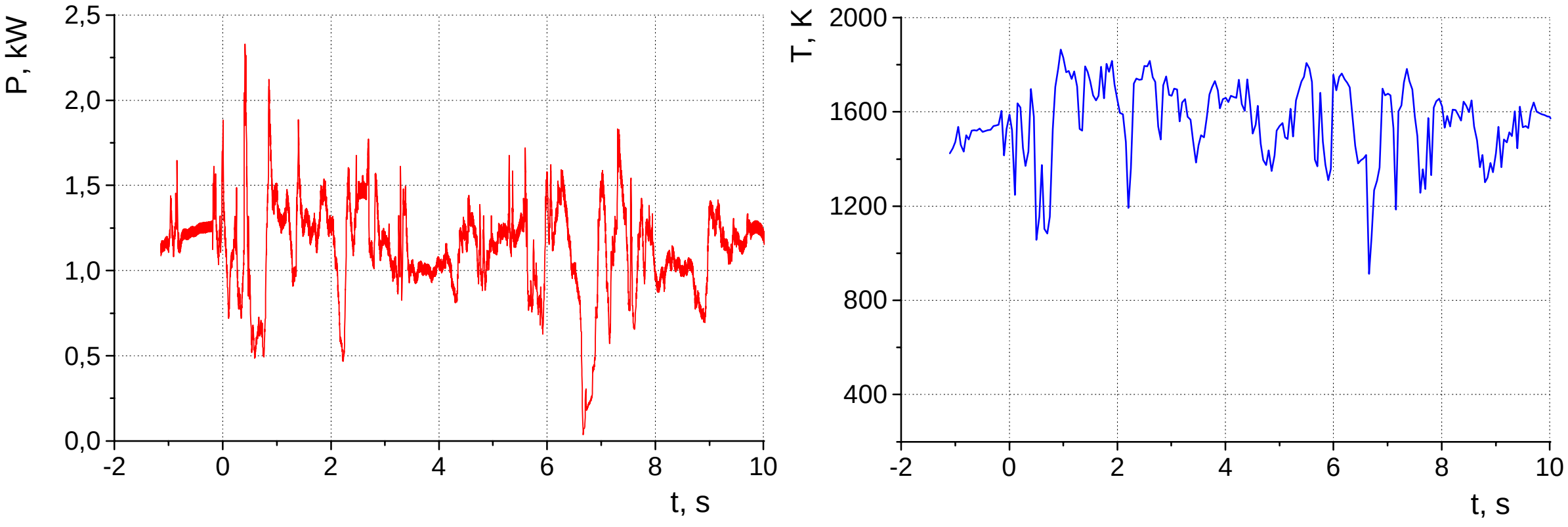

Fig. 2. Flame heat release (a) and maximum temperature in domain (b) at the calculation with $\xi = 5$ and extinguish mixture compound $X_{CO_2} = 4\%$ , $X_{inh} = 2{,}2\%$.

Results of first series of the test calculations are presented on Table 2. Here the carbon dioxide without inhibitor was used as the fire suppressants. The volume fraction of carbon dioxide values in percent with respect to the total gas volume are put on the first row. Distinguishing by dagger cells on the second row correspond to the calculations without extinguishment. Times of fire extinguishment since suppressant mixture incoming to the domain are presented in the other cells on second row. The minimum extinguishing concentration of carbon dioxide turning out 26% appear overestimated as compared with experimental result 20%, which was obtained in [3, 4].

Results of test calculations for two values of empirical constant $\xi$ in expression (11) are presented below. Value $\xi = 7$ was derived in experiment results treatment was used in the first of these series. The fire extinguishment times was obtained in these calculations are presented in Table 3. The volume fractions of carbon dioxide expressed by percent with respect to the total gas volume in the extinguishing mixture incoming to the domain are put in the first columns. The volume fractions of inhibitor are put in the first row. The numerical results in comparison with the experimental data [3, 4] are presented in Fig. 5. The calculations with fire extinguishing are marked by fill points and without ones by empty points. It can be seeing that numerical value of the fire extinguishing inhibitor concentration was underestimated about double time in comparison with the experimental value, which was equal 2.6%. The empirical constant $\xi = 7$ guarantee combustion terminate at 2.0%, so this value mismatch for modeling of the effect of inhibitor on fire by expression (11) with great volume fraction values knowingly. Therefore, value $\xi = 5$ was choosing in other series of calculations. Fire extinguishment times in these calculations are presented in Table 4. The results with comparison on the experimental data are presented in Fig. 6.

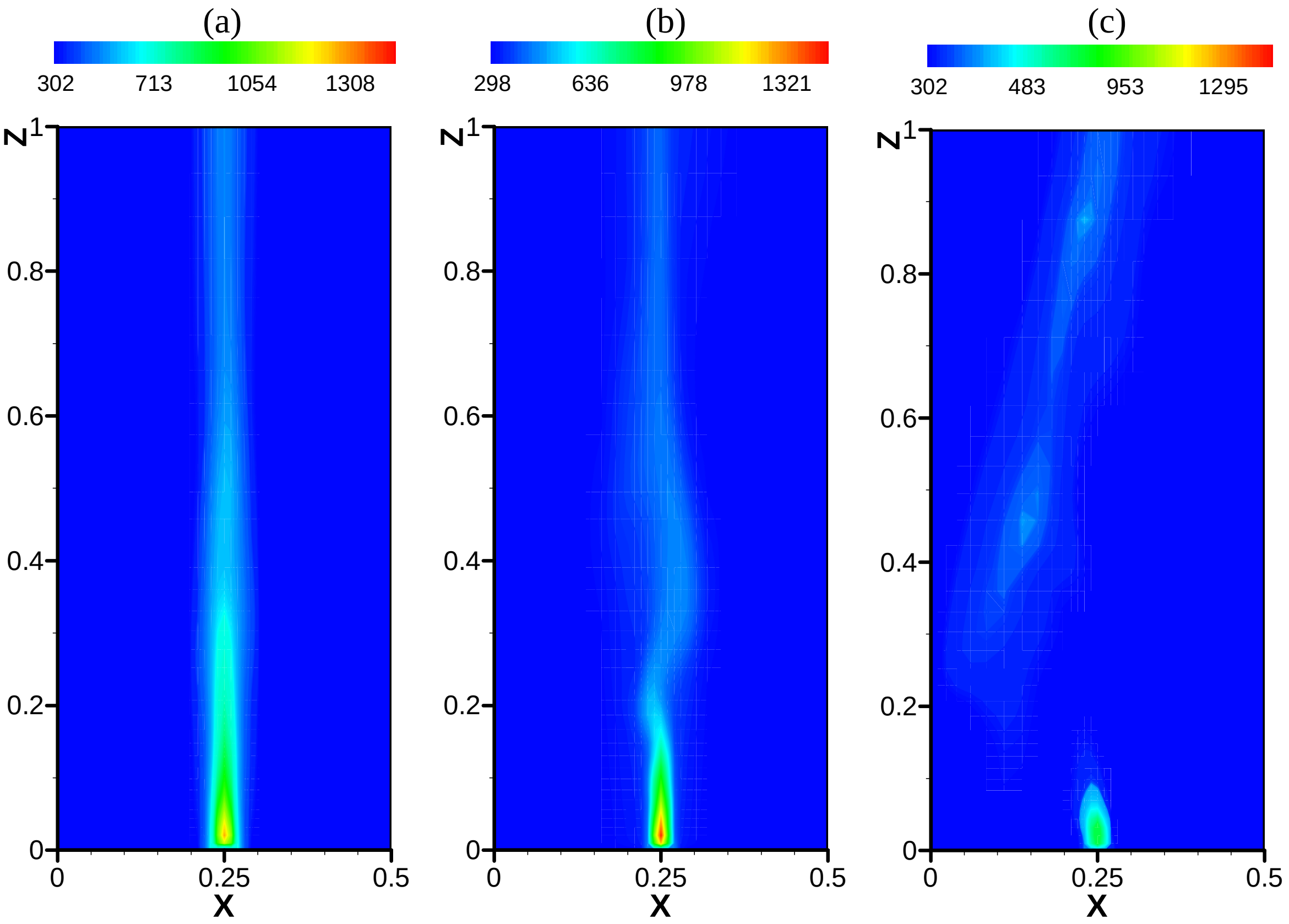

Fig. 3. Temperature distributions for times a) 0 s, b) 1,5 s, c) 4,9 s from outset of extinguishing mixture with the compound $X_{CO_2} = 27\%$ , $X_{inh} = 0\%$ incoming to the domain.

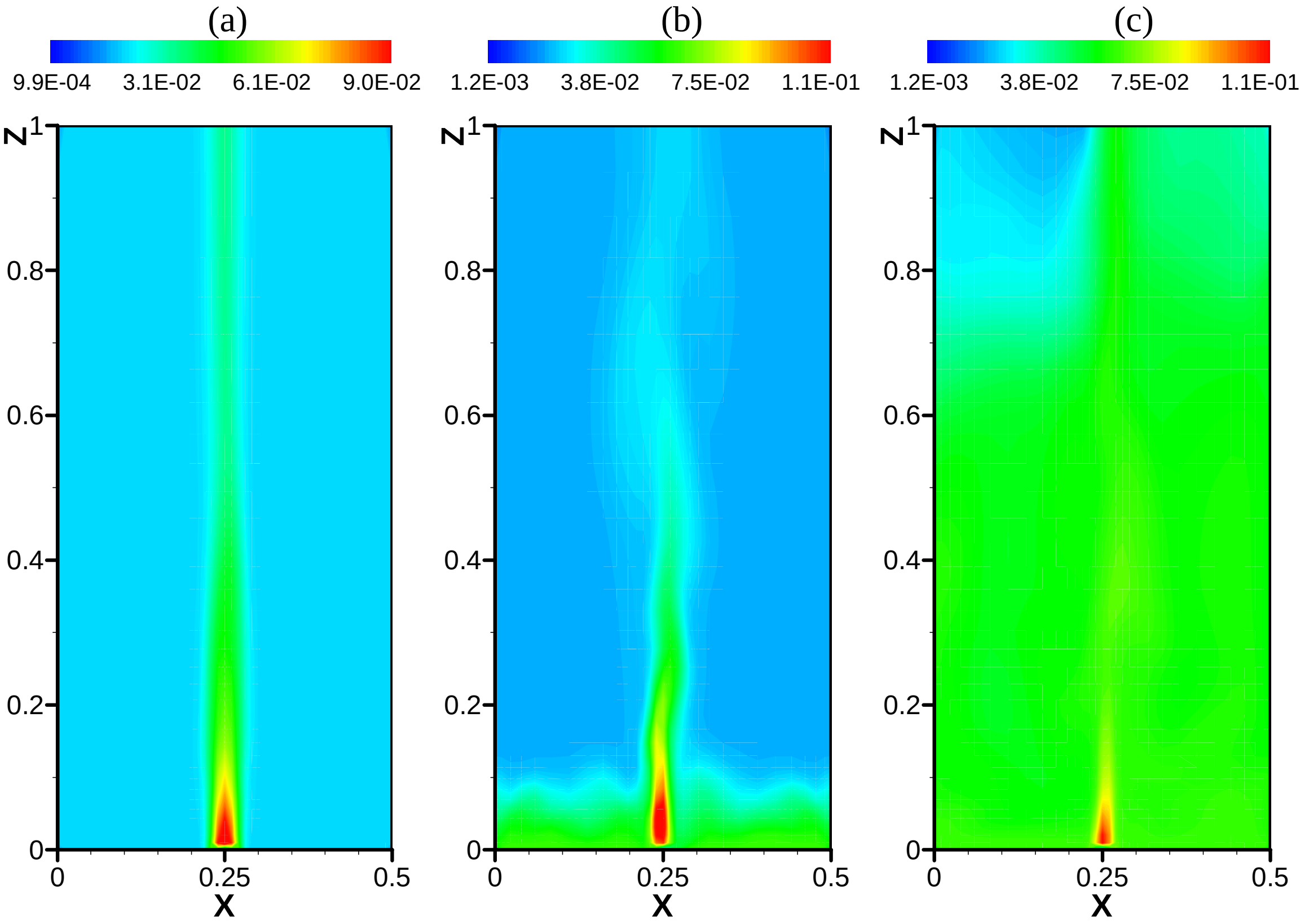

Fig. 4. Carbon dioxide mass fraction distributions for times a) 0 s, b) 0,4 s, c) 7,8 s from outset of extinguishing mixture with the compound $X_{CO_2} = 4\%$ , $X_{inh} = 2{,}2\%$ incoming to the domain for calculation with $\xi = 5$.

Table 3. Flame extinguishments times for $\xi = 7$.

| Inh / $CO_2$ | 0,3 | 0,5 | 0,8 | 0,9 | 1,0 | 1,1 | 1,2 | 1,3 | 1,4 | 1,5 | 1,6 | 1,7 | 2,0 |
|---|---|---|---|---|---|---|---|---|---|---|---|---|---|
| 0 | | | | | | × | × | 5,3 | | 3,9 | | 4,1 | 2,7 |
| 2 | | | | | × | 6,9 | × | | 2,8 | | 2,7 | | |
| 4 | | × | | | 5,3 | × | × | 5,7 | | | | | |
| 6 | | × | × | 9,5 | | | | | | | | | |
| 8 | | | × | 9,1 | | | | | | | | | |
| 12 | | | × | 4,6 | | | | | | | | | |
| 15 | × | 7,2 | 5,4 | | | | | | | | | | |
| 17 | 7,8 | 4,8 | | | | | | | | | | | |

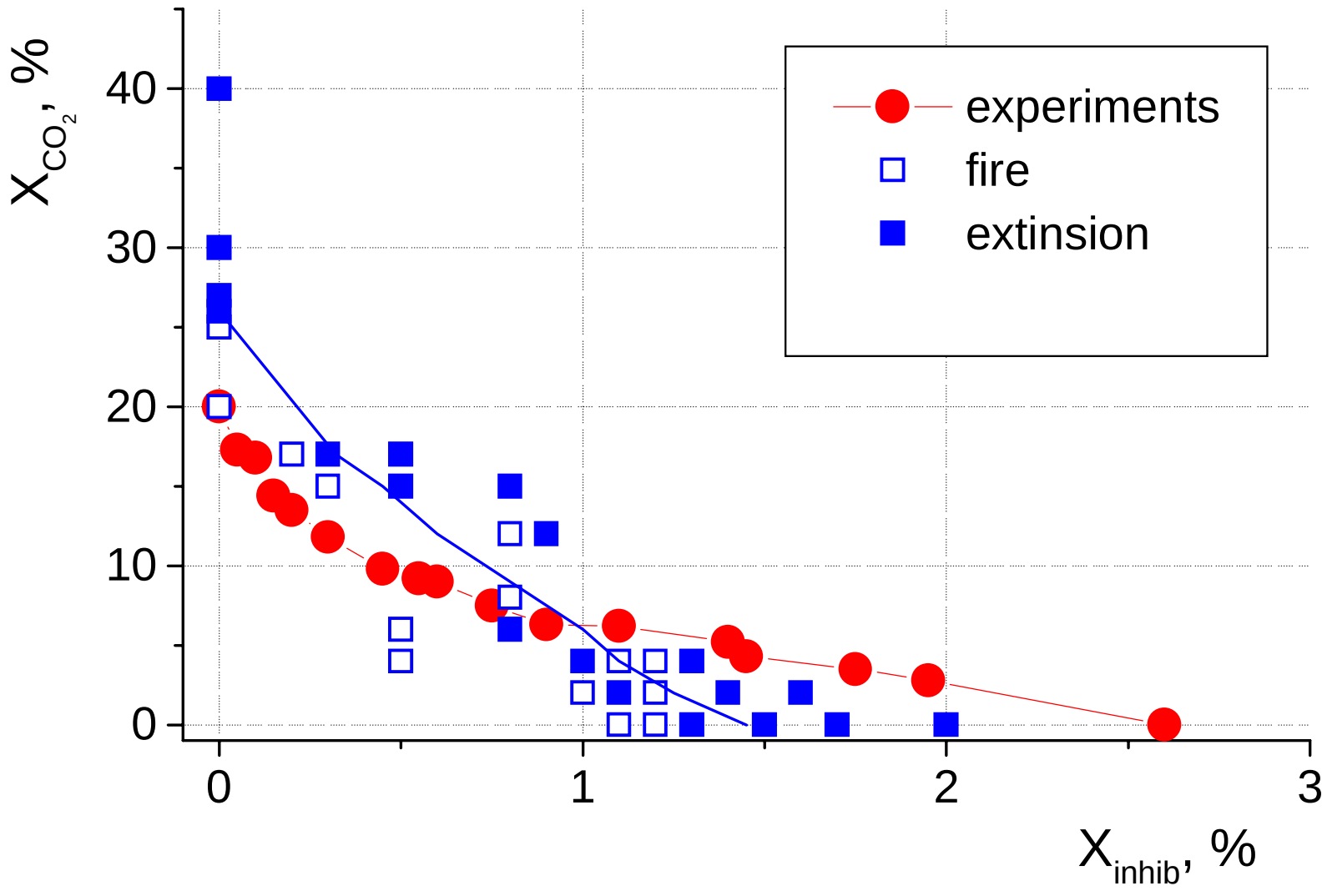

Fig. 5. Results of flame extinguishments for ξ = 7 with the experimental data.

Table 4. Flame extinguishments times for $\xi = 5$.

| Inh<br>$CO_2$ | 0,2 | 0,5 | 0,8 | 1,0 | 1,1 | 1,6 | 1,8 | 2,0 | 2,2 | 2,3 | 2,5 | 2,6 |
|---|---|---|---|---|---|---|---|---|---|---|---|---|
| 0 | | | | | | | | | | | × | 8,5 |
| 2 | | | | | | | | × | × | 8,5 | | |
| 4 | | | | | | | | × | × | 3,9 | | |
| 6 | | | | | | 4,6 | × | 3,9 | 3,4 | | | |
| 8 | × | 5,6 | | × | 5,1 | | | | | | | |
| 12 | | | × | 3,5 | | | | | | | | |
| 15 | | × | | 9 | | | | | | | | |
| 20 | × | 5,4 | | | | | | | | | | |

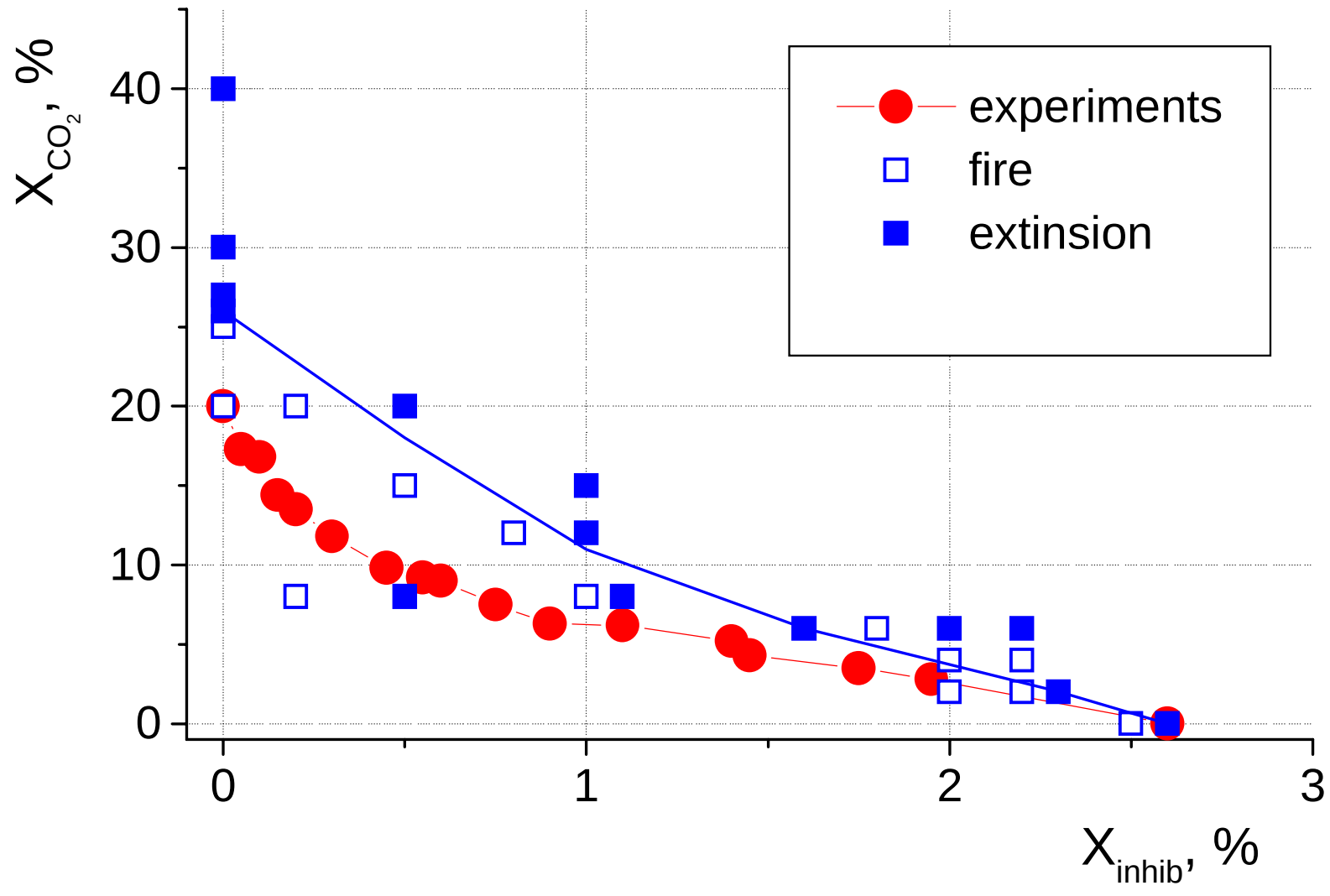

Fig. 6. Results of flame extinguishments for $\xi = 5$ with the experimental data.

Results of both series of calculations with inhibitor permit to estimate the minimum extinguishing concentration of inhibitor, which are drawn in Fig. 5 and Fig. 6 by solid curves. As the result of comparison these diagrams it may be do the conclusion, the empirical expression (11) allow to receive more accurate results with $\xi = 5$ for great volume fraction of inhibitor and with $\xi = 7$ for values less 1%.

## Result of compartment flame extinguishing

The test data were obtaining below were using for the compartment fire extinguishment modeling. The standard compartment is taken into consideration with sizes 2.8×2.8×2.18 m and wall thickness 0.1 m, as was using in [17] (look at Fig. 7). The room with the adiabatic walls and a subspace outside the door opening are included in the computational domain with the sizes 3.5×2.8×2.18 m. There is using grid with the sizes 43×32×23 points. The door opening adjacent to the floor is located on one of the flank side walls along its symmetry axis and its sizes was chosen 0.74 m in width and 1.83 m in high up. The square burner of the size of 0.1×0.1 m is located on the floor in center of the room. The propane flux provides the heat release rate 61.9 kW.

At the beginning the calculation was carried out as long as stationary flow was establishing in the room. Fire extinguishment was doing by pure inhibitor without air and carbon dioxide, which was incoming through a hole with the size of 0.1×0.1 m on the floor in center of the door opening. This position of the fire suppressants source was chosen by following considerations. Ambient air incoming through the lower part of the door opening will flow to the flame directly. Thus, the fire suppressants will suck in the room and will deliver to region with the greatest activity of chemical reaction as soon as possible. Linear velocity of fire suppressants incoming flow was chosen by 1.0 m/s. Use of the test results the empirical coefficient in expression (11) $\xi$ was chosen equal 6.

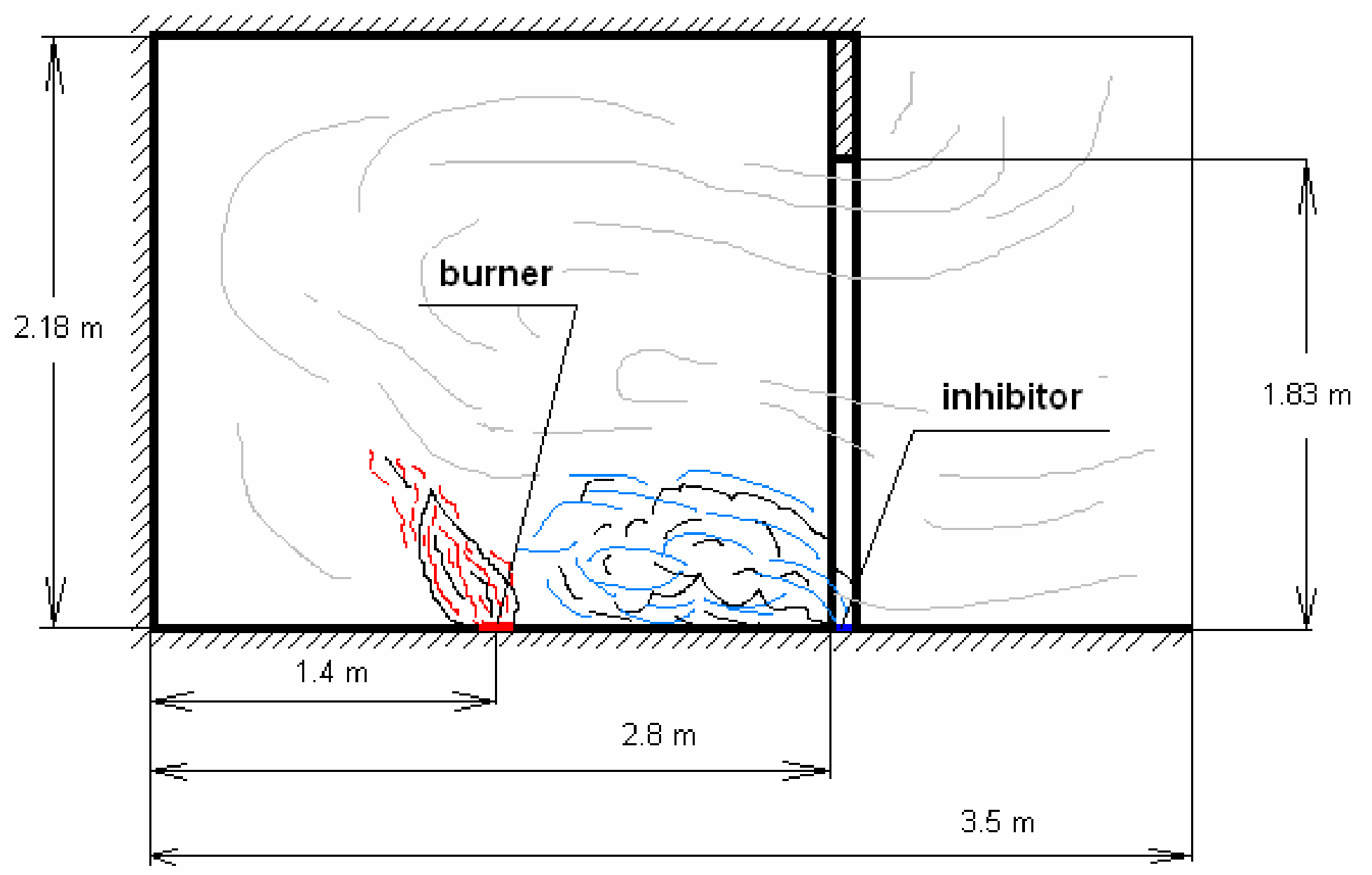

Fig. 7. Computational domain plane for the compartment fire extinguishment modeling.

The temperature and inhibitor volume fraction distributions in the domain vertical symmetry plane at the moment of fire extinguishment are presented on Fig. 8. The temperature distribution was imaged by colors and the inhibitor concentration was described by isolines. Diagrams on Fig. 8 a) corresponds to moment directly before flame extinguishment and on Fig. 8 b) in 1.55 s after it. The time dependence of heat release rate is

a)

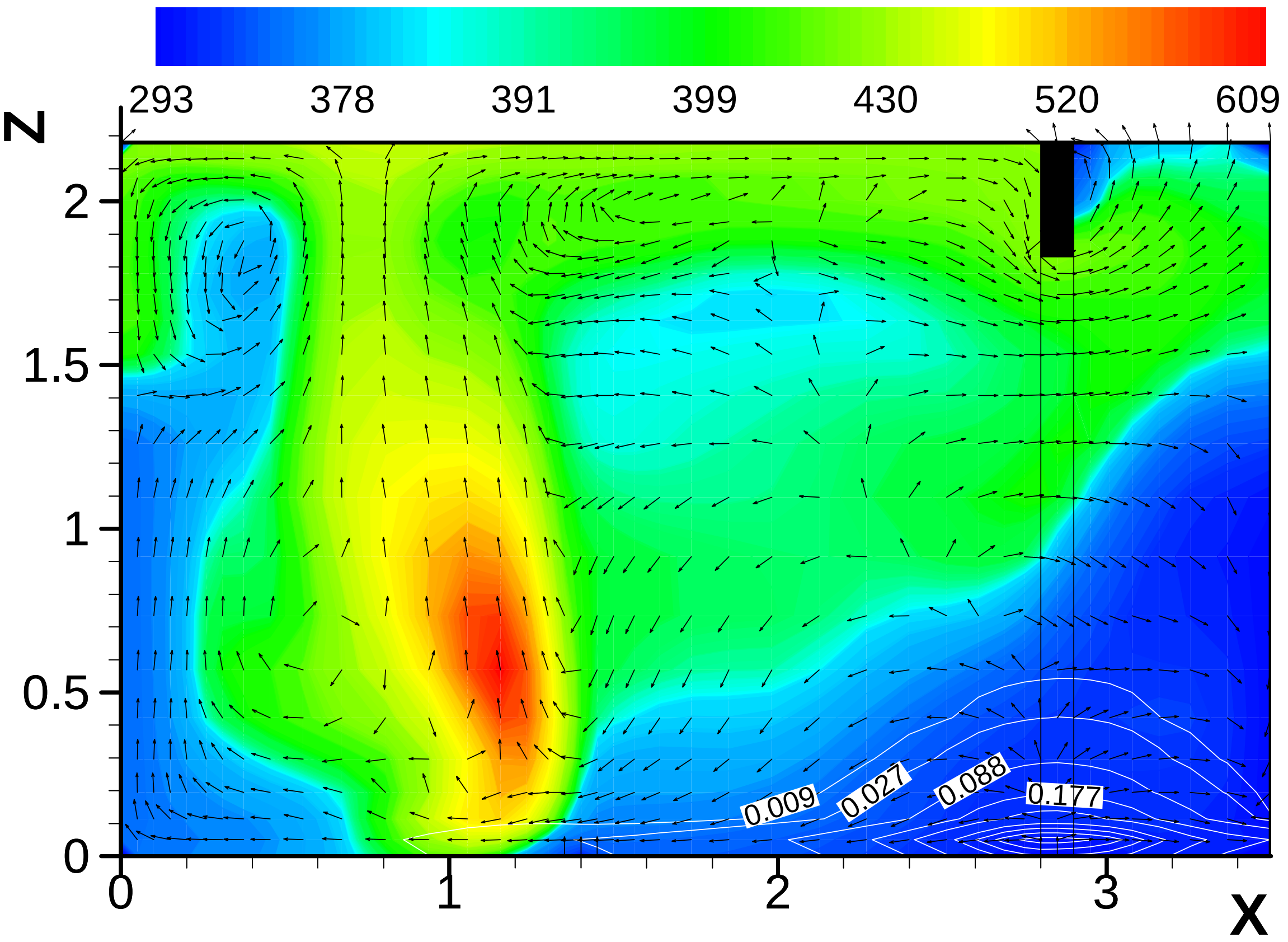

b)

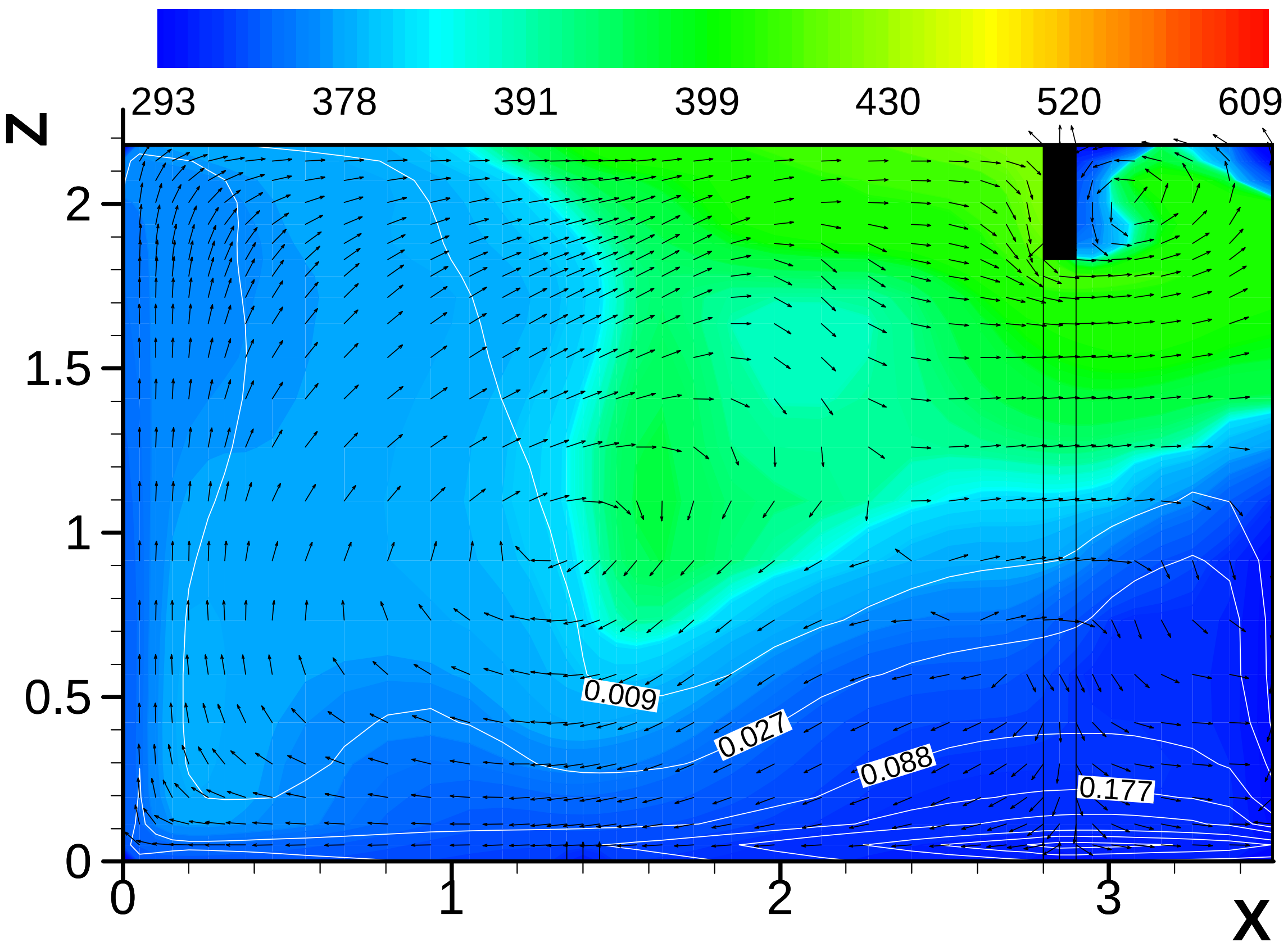

Fig. 8. Temperature and inhibitor volume fraction distributions at the moments a) 11,75 s and b) 13,3 s.

presented on Fig. 9. It can be seen that combustion termination ensued in 11.8 s after beginning of the fire suppressants entry.

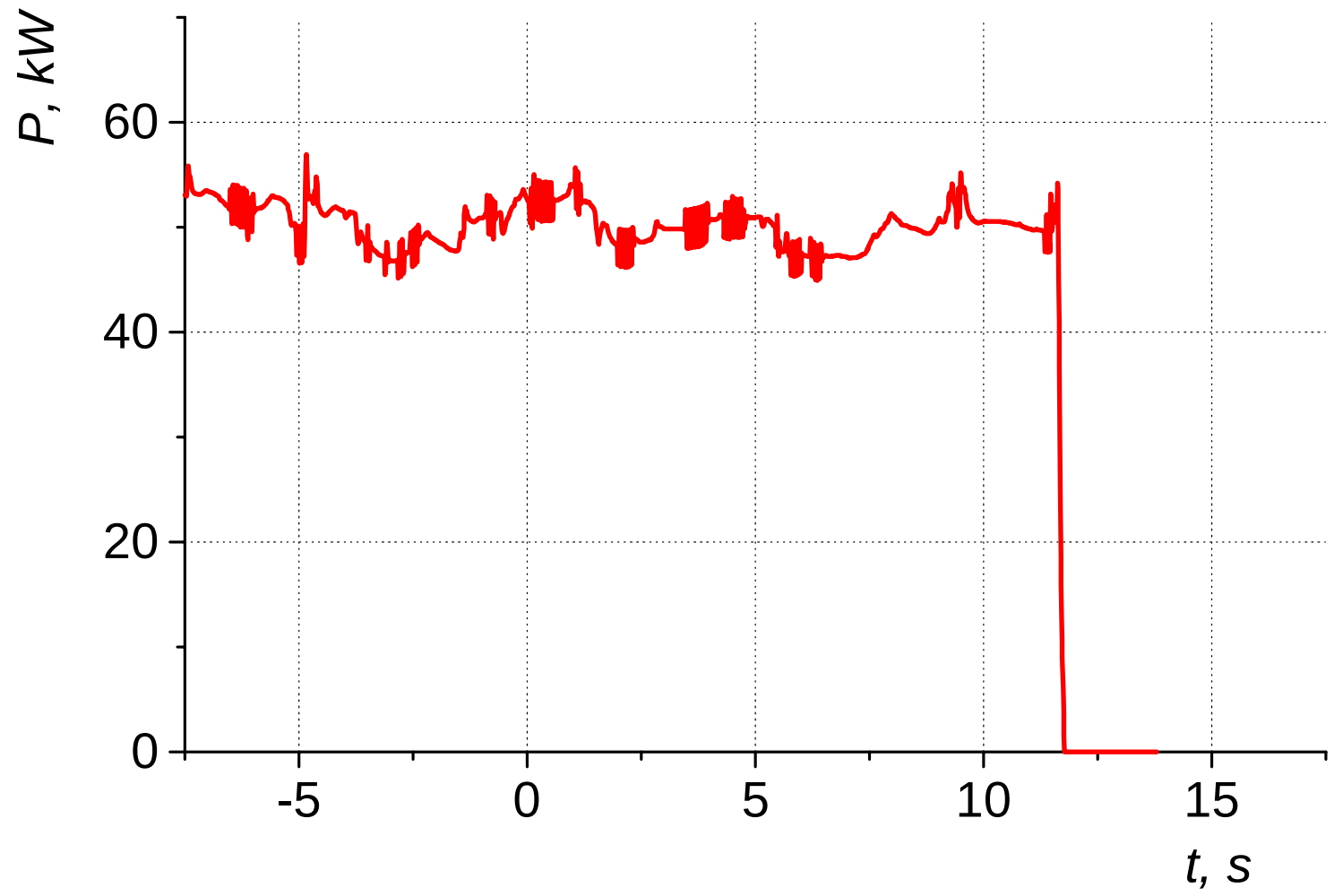

Fig. 9. Flame heat release rate in compartment fire extinguishment calculation.

## Conclusions

Fundamental possibility of numerical modeling of organophosphorous inhibitor $(CF_3CH_2O)_3P$ and its mixtures with carbon dioxide influences on propane combustion method describing below was demonstrated in the calculations. At the same time, comparison of the computational results with experimental data [3, 4] has shown the necessity of empirical expression (11) correction. The overestimated minimum extinguishing concentration of carbon dioxide (without inhibitor) can be attributed to the roughness computational grid and different flame heat release rate in the experiments. Keeping this in mind, we can conclude that the results to agree reasonably with the experimental data.

Compartment fire extinguishment calculation are used to find out the optimum position of the extinguish mixture source.

## Acknowledgments

This work was supported by the INTAS (Grant No. 03-51-4724) and UK Engineering and Physical Sciences Research Council.